\documentclass[twocolumn,showpacs,prl,10pt]{revtex4}
\usepackage[english]{babel}
\usepackage{amsmath,amssymb}
\usepackage{times}
\usepackage{epsfig}

\begin{document}
%\draft

\title {Crossover to non-Fermi-liquid spin dynamics in cuprates} 
\author{J. Bon\v ca$^{1,2}$, P. Prelov\v sek$^{1,2}$, and
I. Sega$^1$} 
\affiliation{$^1$J.\ Stefan Institute, SI-1000 Ljubljana,
Slovenia} 
\affiliation{$^2$Faculty of Mathematics and Physics,
University of Ljubljana, SI-1000 Ljubljana, Slovenia} 
\date{\today}

\begin{abstract}
The antiferromagnetic spin correlation function $S_{\bf Q}$, the
staggered spin susceptibility $\chi_{\bf Q}$ and the energy scale
$\omega_{FL}=S_{\bf Q}/\chi_{\bf Q}$ are studied numerically within
the $t$-$J$ model and the Hubbard model, as relevant to cuprates. It
is shown that $\omega_{FL}$, related to the onset of the
non-Fermi-liquid spin response at $T>\omega_{FL}$, is very low in the
regime below the 'optimum' hole doping $c_h < c_h^* \sim 0.16$, while
it shows a steep increase in the overdoped regime. A quantitative
analysis of NMR spin-spin relaxation-rate $1/T_{2G}$ for various
cuprates reveals a similar behavior, indicating on a sharp, but
continuous, crossover between a Fermi-liquid and a non-Fermi-liquid
behavior as a function of doping.

\end{abstract}

\pacs{71.27.+a, 75.20.-g, 74.72.-h} 
\maketitle 

The understanding of the phase diagram of cuprates continues to
exemplify one of the major theoretical and experimental challenges
\cite{imad}.  Besides superconductivity (SC) and antiferromagnetic
(AFM) ordering, several regimes with distinct electronic properties
have been identified within the normal metallic phase. The behavior of
spin degrees of freedom, which are the subject of this paper, has been
intensively studied using the inelastic neutron scattering (INS)
\cite{regn,kast} and NMR relaxation experiments \cite{bert}. They
clearly reveal that in underdoped cuprates magnetic properties are not
following the usual Fermi-liquid (FL) scenario within the metallic
state above the SC transition $T>T_c$.

Within a normal FL one expects a dynamical spin susceptibility
$\chi_{\bf q}^{\prime\prime}(\omega)$ to be $T$- independent at low
$T,\omega$.  On the contrary, INS results show that ${\bf
q}$-integrated spin susceptibility exhibits in a broad range of
$\omega$ and $T$ an anomalous, but universal behavior
$\chi_L''(\omega) \propto f(\omega/T)$, first established in
La$_{2-x}$Sr$_x$CuO$_4$ (LSCO) at low doping \cite{keim,kast}. This
behavior can be even followed to lowest $T$ in YBaCu$_3$O$_{6+x}$
(YBCO), where $T_c$ has been suppressed by Zn doping \cite{kaku}.  At
the same time, low-energy INS reveals at low $T$ the saturation of the
inverse AFM correlation length $\kappa=1/\xi$, at least in YBCO
\cite{regn} and in LSCO at low doping \cite{keim,kast}. Anomalous
$T$-dependence of $^{63}$Cu NMR spin-lattice relaxation rate $1/T_1$
and of the spin-spin relaxation rate $1/T_{2G}$ in underdoped cuprates
is in general compatible with INS \cite{bert}, in particular $1/(T_1
T) \propto \chi_L''(\omega,T)/\omega |_{\omega \to 0} \propto 1/T$, in
contrast to a $T$-independent value (Korringa law) in a normal FL.

On the other hand, cuprates at optimum doping and, moreover, in the
overdoped regime show a strong reduction of the spin response at low
energies $\omega$. This is evident from the loss of INS intensity in
the normal state (as well as in a weak resonant peak for $T<T_c$) and
low NMR relaxation rates $1/T_1, 1/T_{2G}$. At the same time, NMR
confirms the approach to the normal FL behavior, $1/(T_1T) \sim$
const. and $1/T_{2G} \sim$ const. \cite{bert}. There are other
indications that the normal FL behavior is approached in the overdoped
regime. Recently, the angle-resolved photoemission spectroscopy
(ARPES) on BiSrCaCuO (BSCCO) system gave evidence for the existence of
coherent electronic excitations for $T<T_X$ at higher doping
\cite{kami}, i.e., the FL-like phase is found in the normal state only
in the overdoped regime where $T_X$ shows a steep increase with hole
doping $c_h$.  Intimately related to the onset of the FL-like spin
response is also the observation that in cuprates doped with
nonmagnetic Li and Zn the impurity-induced spin susceptibility varies
as $\propto 1/(T+ T_K)$, i.e., with a Kondo-like behavior with a
characteristic temperature $T_K(c_h)$ \cite{bobr}, where $T_K \sim 0$
in the underdoped regime, whereas it shows a strong increase in the
overdoped regime.

From the point of theoretical understanding, an approach to a FL
behavior in the overdoped regime far from a metal-insulator transition
seems plausible, nevertheless a solid theoretical evidence is still
missing. A crossover from a strange metal to a coherent metal phase
is, e.g., predicted within the slave-boson approach \cite{lee}.
Frequently invoked interpretation is given in terms of the quantum
critical point (QCP) at optimum doping $c_h^*$ (masked, however, at
low $T$ by the SC phase), dividing the FL phase at $c_h>c_h^*$ and a
(singular) non-Fermi-liquid (NFL) metal at $c_h<c_h^*$. While such a
concept is well established in spin systems \cite{sach}, its
application to metallic cuprates is controversial due to the absence
of a critical length scale ( e.g., $\xi(T\to 0)\to \infty$). Low
energy spin dynamics as emerges from INS and in particular from NMR
experiments has been extensively analysed within the phenomenological
theory \cite{mmp}, describing a FL close to an AFM instability.
In the latter approach the spin-fluctuation energy in fact plays the
characteristic FL scale, as discussed furtheron.

The present authors recently showed that an anomalous $\omega/T$
scaling, as observed at low doping, emerges from a general approach to
$\chi_{\bf q}(\omega)$ under a few basic requirements \cite{prel}: a)
collective $\bf Q=(\pi,\pi)$ AFM mode in the normal state is
overdamped $\gamma > \omega_{\bf Q}$, b) equal-time correlations
$S_{\bf Q}=\langle S^z_{-{\bf Q}} S^z_{\bf Q}\rangle$ and the
corresponding inverse correlation length $\tilde \kappa$ are finite
and saturate at low $T$. A nontrivial $\omega_{\bf Q}(T)$ dependence
then follows from the fluctuation - dissipation relation,
\begin{equation}
\frac{1}{\pi}\int_0^\infty d\omega ~{\rm cth}\frac{\omega}{2T}
\chi^{\prime\prime}_{\bf q}(\omega)= S_{\bf q}\, ,
\label{eqsum}
\end{equation}
(note that we use $\hbar=1$ and define $\chi_{\bf q}(\omega)$ in
units of $g^2\mu_B^2$), leading to a $\omega/T$ scaling for $T>
\omega_p \sim \gamma {\rm e}^{-2\zeta}$ where $\zeta \propto
\gamma/\tilde \kappa^2$ \cite{prel}.  Within such an approach it is
natural that $\omega_p>0$ is finite within the whole paramagnetic
regime. Nevertheless, due to strong dependence on parameters, in
particular on $\zeta$, $\omega_p(c_h)$ can show quite a sharp
crossover from very small values in the underdoped regime to a large
increase in overdoped systems, consistent with experimental
indications.

In this paper we present numerical results for the doping dependence
of the FL scale $\omega_{FL}$ within models relevant to cuprates,
i.e., the planar $t$-$J$ model and the Hubbard model. We furthermore
compare these quantities with the ones extracted directly from
NMR-relaxation and INS experiments on cuprates.  One possibility is to
get $\omega_{FL}$ from the full $T$-dependence of various magnetic
quantities, in particular from static $\chi_{\bf Q}(T)$ and $S_{\bf
Q}(T)$. It is evident that in the NFL regime $T>\omega_{FL}$ a
relation follows from Eq.~(\ref{eqsum}),
\begin{equation}
\frac{S_{\bf Q}}{T\chi_{\bf Q}} = \Bigl[ 1 - \frac{\Delta}{S_{\bf Q}}
\Bigl]^{-1}, 
\label{eqcl}
\end{equation}
which evolves into the `classical' relation for $\Delta \ll S_{\bf
Q}$. Note that $\Delta(T)$ arises from Eq.~(\ref{eqsum}) as the
integral over the large-$\omega$ tail $\chi_{\bf Q}''(\omega>T)$. We
are interested in the low-$T$ regime in the paramagnetic phase where
$S_{\bf Q}(T)$ already saturates. The saturation is quite evident from
the numerical analysis of various models
\cite{sing,imad}. Eq.~(\ref{eqcl}) indicates that even constant
$S_{\bf Q}$ can be compatible with strongly $T$-dependent $\chi_{\bf
Q}(T)$ which seems to be the essence of the NFL regime in cuprates. In
contrast, one expects a finite $\chi_{\bf Q}(T\to
0)$ within the FL regime. 

The characteristic energy scale of spin fluctuations is given by
$\omega_{FL}(T)=S_{\bf Q}/\chi_{\bf Q}(T)$ with the corresponding
$T=0$ limit $\omega_{FL}(0)$. The latter can be calculated from $T=0$
numerical results.  Note that $\omega_{FL}(0)=\langle \omega \rangle$
is just the first frequency moment of the shape function $\chi_{\bf
Q}''(\omega,T=0)/\omega$ for $\omega>0$. On the other hand, we extract
$\omega_{FL}$ also from experiments, in particular from NMR $1/T_{2G}$
relaxation data, which give rather straightforward information on
$\chi_{\bf Q}(T)$.
 
Let us first consider the $t$-$J$ model,
\begin{equation}
H=-t\sum_{\langle ij\rangle s} \tilde{c}^\dagger_{js}\tilde{c}_{is}
+J\sum_{\langle ij\rangle}({\bf S}_i\cdot {\bf S}_j-\frac{1}{4}
n_in_j) \, , \label{eqtj}
\end{equation}
with the nearest neighbor hopping on a square lattice, which we analyse
for $J/t =0.3$, as relevant for cuprates (for comparison with cuprates
we use also $t \sim 400$ meV). Results for $S_{\bf Q}(T)$ and
$\chi_{\bf Q}(T)$ are evaluated using the finite-$T$ Lanczos method
(FTLM) \cite{jakl}. In this way we analyse systems with $N=18$ sites
for arbitrary hole doping $c_h=N_h/N$, and with $c_h \leq 3/20$ for
$N=20$.  It should be also noted that FTLM results are rather
insensitive to finite-size effects for $T>T_{fs}$, whereby for systems
considered $T_{fs} \sim 0.1~t$ \cite{jakl}.

In Fig.~1 we present results for $\tilde \chi=4T\chi_{\bf Q}$ as a
function of $c_h$ for various $T>T_{fs}$. Note that the limiting value
within the $t$-$J$ model is $\tilde \chi(T\to \infty) = 1-c_h$. Two
distinct regimes become immediately evident from Fig.~1. The crossing
of curves $\tilde \chi(c_h)$ with different $T$ can be used as the
definition of the 'optimum' doping $c_h^* \sim 0.16$, whereby it is
indicative that the same value is obtained analysing cuprates with
highest $T_c$ \cite{coop}.  In the underdoped regime $\tilde \chi$
increases by lowering $T$ (down to reachable $T \sim T_{fs}$) and
appears to saturate to the NFL behavior, Eq.~(\ref{eqcl}), consistent
with the anomalous $\omega/T$ scaling \cite{prel}. On the other hand,
at $c_h>c_h^*$ the tendency of $\tilde \chi(T)$ is opposite. I.e.,
$\chi_{\bf Q}(T)$ saturates for $T <J$, indicating a 'normal' FL
behavior. If $\tilde \chi(c_h)$ curves would, even for lowest $T$,
indeed cross at $c_h=c_h^*$, we would have been dealing with a
singularity resembling a QCP with diverging $\chi_{\bf Q}(T\to 0)
\propto 1/T$. Moreover $\chi_{\bf Q}(T\to 0)$ would be divergent in
the whole regime $c_h<c_h^*$. Although present results cannot exclude
this possibility, the deviation visible at lowest $T=0.1~t$ is more in
accord with a crossover between FL and NFL regimes.

\begin{figure}[htb]
\centering
\epsfig{file=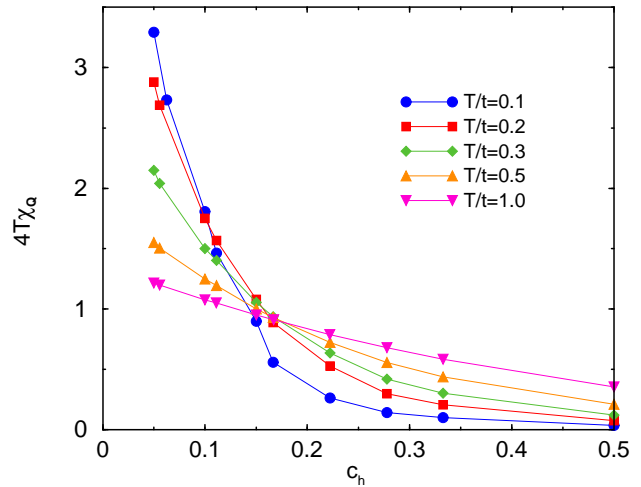,width=65mm,angle=-90}
\caption{(color online) AFM susceptibility $4T\chi_{\bf Q}$ vs. doping
within the $t$-$J$ model for various temperatures $T/t$.}
\label{fig1}
\end{figure}

In Fig.~2 we show corresponding FTLM results for $\omega_{FL}(c_h)$ at
various $T\leq J$. Note that in this regime $S_{\bf Q}(T)$ is
essentially $T$-independent, and the values agree very well with the
$T=0$ results obtained via the usual Lanczos technique for the ground
state (g.s.). The latter approach allows the calculation of $S_{\bf
Q}(T=0)$ and $\chi_{\bf Q}(T=0)$ also for somewhat larger systems,
i.e., for $N=20$ at all $N_h$ and for $N=26$ at $N_h\leq 2$. $T=0$
results for $S_{\bf Q}$ are shown in the inset of Fig.~2 and overall
follow surprisingly well the linear variation $1/S_{\bf Q}=K c_h$ with
$K \sim 15$. In contrast to $S_{\bf Q}(c_h)$, the FL scale
$\omega_{FL}$ reveals a nonuniform variation with doping. Again, for
$c_h>c_h^*$ $\omega_{FL}$ is already $T$-independent for $T<J$, or at
least approaching finite $\omega_{FL}(0)$. In the regime $c_h<c_h^*$
$\omega_{FL}(T)$ is decreasing with $T$, so that we can establish only
an upper bound for $\omega_{FL}$. In the same Fig.~2 we plot also
results for $\omega_{FL}(0)$, evaluated directly via the $T=0$
calculation for largest available systems. In the overdoped regime the
general agreement with the FTLM is evident. As expected, in the
underdoped region obtained $\omega_{FL}(0)$ seem to be consistently
smaller that $\omega_{FL}(T>0)$ values, whereby a decrease of
$\omega_{FL}(0)$ with system size is also observed (e.g., values
obtained for $N=26$ systems are smaller than those for $N=18,20$). So
we can summarize results in Fig.~2 as follows: a) in the overdoped
regime $\omega_{FL} \sim \alpha (c_h-c_{h0})$ with $c_{h0} \sim 0.12$
and a large slope $\alpha \sim 3.9t \sim 1.6$ eV, b) in the underdoped
regime our results seem to indicate on a smooth crossover to very
small $\omega_{FL} \ll J$.

\begin{figure}[htb]
\centering
\epsfig{file=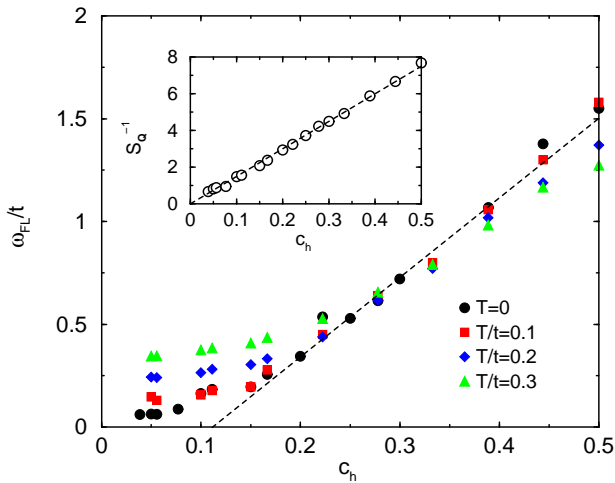,width=65mm,angle=-90}
\caption{(color online) FL scale $\omega_{FL}/t$ vs. $c_h$, obtained
for the $t$-$J$ model using the FTLM for $T>0$ and the usual Lanczos
method for $T=0$. The inset shows $T=0$ results for $1/S_{\bf Q}$
vs. $c_h$. Thin lines are guide to the eye only.}
\label{fig2}
\end{figure}

The alternative relevant model is the Hubbard model on a square
lattice,
\begin{equation}
H=-t\sum_{\langle ij\rangle s}( c_{is}^\dagger
c_{js}^{\phantom{\dagger}}+\text{H.c.})+ U\sum_{i}
n_{i\uparrow}n_{i\downarrow}, \label{eqhub}
\end{equation}
which in the case of strong Coulomb repulsion $U \gg t$ and close to
half-filling maps onto the $t$-$J$ model with $J=4t^2/U$. We calculate
$S_{\bf Q}$ in the g.s. as a function of hole doping $c_h$ within the
Hubbard model on a square lattice and at $U=8~t$ using the
constrained-path quantum Monte Carlo method (CPMC) \cite{zhang}. In
this method, the ground state wave function is projected from a known
initial wave function by a branching random walk in the overcomplete
space of Slater determinants. Since the method is most efficient in
the closed-shell cases, we extend our calculations to various tilted
square lattices where the number of sites $N$ is ranging between
$N=34$ to $N=164$.  The susceptibility $\chi_{\bf Q}= \partial m_{\bf
Q}/\partial B_{\bf Q} $ is calculated by computing sublattice
magnetization $m_{\bf Q}$ induced by a small staggered magnetic field
$B_{\bf Q}$.

Our results for $1/S_{\bf Q}$ again reveal a linear variation $\sim K
c_h$ with $K \sim 14$. Such results are in qualitative agreement with
previous QMC calculations for $U/t=4$ \cite{imad}, where in the latter
case $K \sim 14.3$. In Fig.~3 we present corresponding
$\omega_{FL}(0)$. The qualitative behavior of $\omega_{FL}$ is very
similar to the result within the $t$-$J$ model, Fig.~2. In the
overdoped regime one can again approximate the variation of
$\omega_{FL}$ as linear, with $c_{h0}\sim 0.1$ and $\alpha \sim
4.8~t$, while for $c_h<c_{h0}$ $\omega_{FL}$ becomes very
small. Altogether, obtained $\omega_{FL}$ do not differ much from that
within the $t$-$J$ model, in spite of plausibly weaker correlations
within the Hubbard model for $U=8~t$ \cite{bonc}. In Fig.~3 we display
also the corresponding free fermion result. We notice that on
approaching the empty band $c_e =1-c_h\to 0$ both curves
converge. However, close to half-filling there is a huge qualitative
difference.

\begin{figure}[htb]
\centering
\epsfig{file=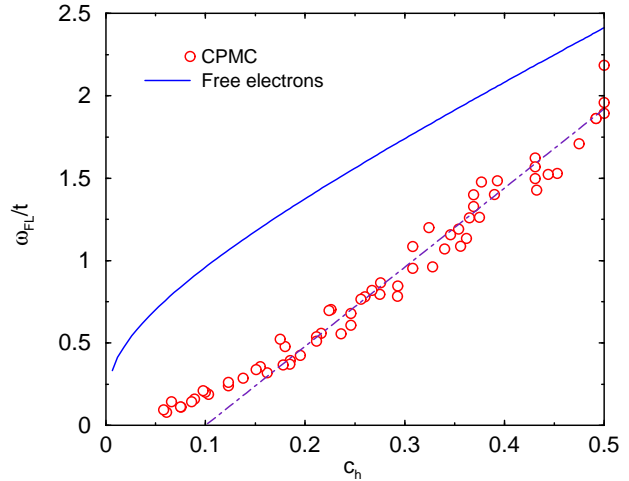,width=65mm,angle=-90}
\caption{(color online) FL scale $\omega_{FL}/t$ vs. doping $c_h$, as
obtained via the CPMC method for the Hubbard model with $U/t=8$, where
the dashed line is a guide to the eye.}
\label{fig3}
\end{figure}

Let us finally estimate $\chi_{\bf Q}(T)$ and consequently
$\omega_{FL}$ directly from experiments on cuprates. Within the normal
state we use the results for the NMR spin-spin relaxation time
$T_{2G}$, obtained from the $^{63}$Cu spin-echo decay, related to
static $\chi_{\bf q}$ as \cite{mmp},
\begin{equation}
\frac{1}{T_{2G}^2}=\frac{0.69}{8} \Bigl[\frac{1}{N}\sum_{\bf q}
 (F({\bf q}) \chi_{\bf q})^2 - \bigl( \frac{1}{N} \sum_{\bf q} F({\bf
 q}) \chi_{\bf q} \bigr)^2 \Bigr]. \label{eqt2}
\end{equation}
Assuming that $\chi_{\bf q}$ is peaked at commensurate ${\bf q}={\bf
Q}$ and can be described by a Lorentzian form $\chi_{\bf q}=\chi_{\bf
Q} \kappa^2/[({\bf q}-{\bf Q})^2+ \kappa^2]$ (e.g., consistent with
INS in YBCO) with $\kappa \ll \pi$, the second term in
Eq.~(\ref{eqt2}) can be neglected and the form factor replaced by
$F({\bf Q})$. This leads to the relation $1/T_{2G} \sim 0.083 \kappa
F({\bf Q}) \chi_{\bf Q}$. $1/T_{2G}$ relaxation rates have been
measured and summarized in Ref.~\cite{bert}, i.e., from underdoped to
optimally doped YBCO with $0.63<x<1$, underdoped YBa$_2$Cu$_4$O$_8$,
nearly optimum doped Tl$_2$Ba$_2$Ca$_2$Cu$_3$O$_{10}$ (Tl-2223) and
the overdoped Tl$_2$Ba$_2$CuO$_{6+\delta}$ (Tl-2201), whereby the
normalization with corresponding $F({\bf Q})$ has been already taken
into account (see Fig.~8b in Ref.~\cite{bert}). Note that $\kappa$
relevant to $\chi_{\bf q}$ is the one appropriate for low-$\omega$
spin dynamics, as measured directly by INS (plausibly $\kappa < \tilde
\kappa$). For YBCO $\kappa(x)$ has been summarized in
Ref.~\cite{bala}. For cuprates considered here appropriate hole
concentrations $c_h$ have been estimated in Ref.~\cite{coop}. Assuming
a continuous variation of $\kappa(c_h)$ we determine also $\kappa$ for
YBa$_2$Cu$_4$O$_8$, Tl-2223 and Tl-2201 (for the latter we take
$\kappa=1.2/a_0$), not available experimentally. In this way, we
evaluate $\chi_{\bf Q}(T)$. Equal-time correlations $S_{\bf Q}$ are so
far not directly accessible by INS. As shown before they are nearly
model independent, so we assume here the $t$-$J$ model results to
finally extract corresponding $\omega_{FL}(T)$ as presented in Fig.~4
for various cuprates.

\begin{figure}[htb]
\centering
\epsfig{file=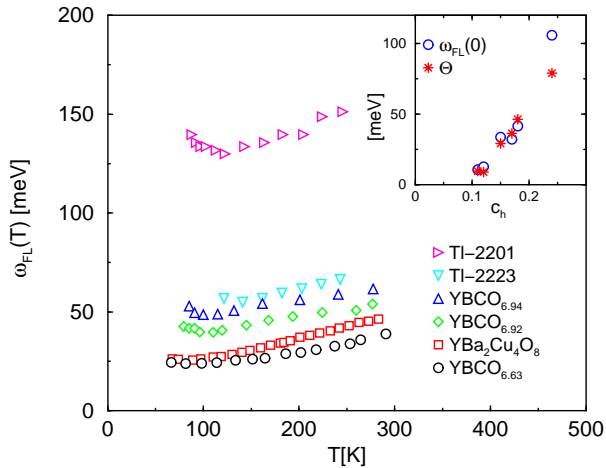,width=65mm,angle=-90}
\caption{(color online) $\omega_{FL}$ vs. $T$, as evaluated from the
NMR relaxation rate $1/T_{2G}$ \cite{bert} and the INS width $\kappa$
\cite{bala} for various cuprates. The inset shows the extrapolated
scales $\omega_{FL}(0)$ and $\Theta$ vs. doping $c_h$.}
\label{fig4}
\end{figure}

For $T$ well above $T_c$, $\chi_{\bf Q}(T)$ extracted from $1/T_{2G}$
follows the Curie-Weiss behavior, i.e., $\chi_{\bf Q}(T) \propto 1/(T+
\Theta)$. Such a behavior emerges also within our analytical approach
in the scaling regime \cite{prel}. Hence, for $T>150K$, we can well
parametrize $\omega_{FL}(T)=\omega_{FL}(0)(1+T/\Theta)$ and present in
the inset of Fig.~4 the doping dependence of $\omega_{FL}(0)$ and
$\Theta$. It is evident that both $\omega_{FL}(0)$ and $\Theta$ reveal
a similar behavior, which qualitatively and to some extent even
quantitatively follow our result within the $t$-$J$ and Hubbard
models. In particular, there is a clear change of scale between the
underdoped and overdoped cuprates.

In conclusion we present evidence, based both on numerical results within
$t$-$J$ and Hubbard models as well as on the analyses of NMR and INS
experimental data on cuprates, that the FL scale $\omega_{FL}$
exhibits a rather sharp crossover between a steep increase in the
overdoped regime and very low $\omega_{FL} \ll J$ in the underdoped
regime for $c_h<c_h^*$. Note that in the latter regime within cuprates
one can easily reach values $\omega_{FL}(0)$ smaller than $T_c$. This
can explain why anomalous NFL scaling of the spin response as well as
of other quantities is observed throughout the normal phase at
$T>T_c$.  On the other hand, the transition to the normal FL is quite
abrupt in the overdoped regime, at least with respect to the spin
response discussed here.

Our results are well in agreement with other experimental evidence for
the existence of transition to the FL behavior in cuprates. The FL
scale $T_X$, as revealed by recent ARPES experiments on BSCCO
\cite{kami}, in particular its doping dependence $T_X(c_h)$ in the
overdoped regime, is close to our results for $\omega_{FL}(c_h)$ with
an extrapolated $c_{h0} \sim 0.1$.  Similar doping dependent scale
$T_K$, analogous to a Kondo scale in metals, arises from the analysis
of the local-moment susceptibilities in YBCO with in-plane nonmagnetic
Li and Zn impurities \cite{bobr}. Experiments show an abrupt and steep
increase of $T_K$ on approaching the optimum doping. It is plausible
that the impurity-induced uniform susceptibility is related to local
$\chi_L$ (and to staggered $\chi_{\bf Q}^0$) in an uniform system,
hence $T_K$ seems to be related to $\Theta$. Needless to say such a
relation requires a theoretical justification.

Still, our numerical results cannot exclude the possibility of the
existence of a QCP. From our analysis, the latter can be present at
the point where $\omega_{FL}(0)$ vanishes on approaching from the
overdoped side, i.e., in our model systems at $c_h \sim c_{h0} <
c_h^*$. An analogous interpretation might follow also from
experimental values in Fig.~4, as well as from results on the Kondo
temperature $T_K(c_h)$ \cite{bobr}. However, the main obstacle to such
a scenario is that there is no evidence for an ordered AFM phase for
$c_h < c_{h0}$, neither from calculated $S_{\bf Q}$ within the $t$-$J$
and Hubbard models nor from experiments.  As our analysis shows
$\omega_{FL}(0)$ remains finite throughout the normal phase at all
dopings $c_h$ down to the onset of the ordered AFM phase at
$c_h^{AFM}<c_{h0}$. The experimental distinction between the QCP and
the present crossover scenario is that in principle $\omega_{FL}(0)>0$
in the normal phase even in the heavily underdoped regime, hence one
should be able to detect this experimentally by suppressing the SC
phase, e.g., as investigated with INS on YBCO system
\cite{kaku}. However, the theory \cite{prel} reveals that
$\omega_{FL}(0)\propto \omega_p$ can be extremely small in the
underdoped regime.

%This work was sponsored by the Slovene Ministry of Education, Science
%and Sports under grant P1-0044.

\end{document}